\documentclass[10pt, compress]{article}
\usepackage[utf8]{inputenc}
\usepackage[T1]{fontenc}
\usepackage{fixltx2e}
\usepackage{graphicx}
\usepackage{longtable}
\usepackage{float}
\usepackage{wrapfig}
\usepackage{rotating}
\usepackage[normalem]{ulem}
\usepackage{amsmath}
\usepackage{textcomp}
\usepackage{marvosym}
\usepackage{wasysym}
\usepackage{amssymb}
\usepackage{hyperref}
\usepackage{listings}
\usepackage{amsmath}
\usepackage{todonotes}
\usepackage{subcaption}
\usepackage{amsthm}
\usepackage{dirtytalk}
\usepackage{natbib}
\theoremstyle{definition}

\theoremstyle{plain}

\author{\textit{Stefano Bennati} \\ \small[\url{stefano.bennati@gess.ethz.ch}] \\ Professorship of Computational Social Science \\ \small{ETHZ, Clausiusstrasse 50, 8092, Zürich Switzerland}}
\date{\today}
\title{On the Role of Collective Sensing and Evolution in Group Formation.}
\begin{document}

\maketitle
\makeatletter
\def\IfClass#1#2#3{\@ifundefined{opt@#1.cls}{#3}{#2}}
\makeatother

\begin{abstract}
Collective sensing is an emergent phenomenon which enables individuals to estimate a hidden property of the environment through the observation of social interactions.
Previous work on collective sensing shows that gregarious individuals obtain an evolutionary advantage by exploiting collective sensing when competing against solitary individuals.
This work addresses the question of whether collective sensing allows for the emergence of groups from a population of individuals without predetermined behaviors.
It is assumed that group membership does not lessen competition on the limited resources in the environment, e.g. groups do not improve foraging efficiency.
Experiments are run in an agent-based evolutionary model of a foraging task, where the fitness of the agents depends on their foraging strategy.
The foraging strategy of agents is determined by a neural network, which does not require explicit modeling of the environment and of the interactions between agents.

Experiments demonstrate that gregarious behavior is not the evolutionary-fittest strategy if resources are abundant, thus invalidating previous findings in a specific region of the parameter space.
  In other words, resource scarcity makes gregarious behavior so valuable as to make up for the increased competition over the few available resources.
  Furthermore, it is shown that a population of solitary agents can evolve gregarious behavior in response to a sudden scarcity of resources, thus individuating a possible mechanism that leads to gregarious behavior in nature.
  The evolutionary process operates on the whole parameter space of the neural networks, hence these behaviors are selected among an unconstrained set of behavioral models.
\end{abstract}
\pagebreak

\section{Introduction}

Group dynamics is a topic of interest in both thefield of sociology  \citep{forsyth2009group,lewin1947frontiers,hamilton2007complex,perc2013evolutionary} and that of biology \citep{krause2002living,giraldeau2000social,couzin2002collective,nowak2010evolutionary}.
Several factors are considered to support the existence of groups:
social hunters see their feeding efficiency increase as groups can prey on larger animals \citep{schoener71_theor_feedin_strat,krebs73_social_learn_signif_mixed_species} and can resist the competition of other groups \citep{puurtinen09_between_group_compet_human_cooper}.
Groups also come with antipredatory advantages as individuals see their chance of survival increase \citep{bertram78_livin_group,elgar1989predator}.
Moreover, close contact with other individuals also increases reproductive opportunities \citep{bertram78_livin_group} and information exchange \citep{valone89_group_forag_public_infor_patch_estim,kurland85_optim_forag_homin_evolut}.

The question that motivates this work is: can collective sensing lead to the emergence of groups in a population of individualistic agents, without making any assumption about the individual behavior?
Collective sensing is an emergent phenomenon present in nature  \citep{berdahl13_emerg_sensin_compl_envir_by}, and it refers to the ability of a group to sense what is beyond the capabilities of the individual.
The additional information about the environment, conveyed through collective behavior, provides individuals with an evolutionary advantage  \citep{elgar1989predator,seeley2009wisdom,hein15_evolut_distr_sensin_collec_comput_animal_popul,torney11_signal_evolut_cooper_forag_dynam_envir,bhattacharya14_collec_forag_heter_lands}.
Information plays a key role in nature: it enabled the evolution of complexity in nature \citep{szathmary95_major_evolut_trans} and it shapes individual behavior \citep{vergassola07_as_strat_searc_without_gradien}, group behavior \citep{skyrms10_signal} and collective intelligence \citep{garnier07_biolog_princ_swarm_intel}.
This paper argues that information can be responsible for the existence of groups.
The information considered in this work is about food location, but it is also suitable for other interpretations, e.g. mating opportunities, suitable nest locations \citep{berdahl13_emerg_sensin_compl_envir_by}.
For example, bird assemblages, e.g. communal roosts, are believed to have evolved as information exchanges \citep{ward1973importance}.
The approach taken in this work is based on simulation and is general across species.

An agent-based simulation environment is used to model a population of agents which perform a foraging task in a patchy environment with hidden resources.
Agents compete for the same limited resources and their fitness depends on their foraging strategy.
Two foraging strategies are compared: random walk, an individualistic strategy that ignores other agents, and gregariousness, which attracts agents towards crowded areas.
Experiments show that gregarious agents have an evolutionary advantage over random walkers if food sources are rare, as it increases the foraging competition between agents, but collective sensing makes up for this disadvantage by increasing the efficiency of finding resources.
  The success of gregarious agents is compromised for high population densities as larger groups increase the rate of resource consumption, thus making exploration more successful than exploitation of known resources.
Moreover, it has been shown that gregariousness can emerge from a population of individualistic agents as an evolutionary response to a reduction in resource availability (see also \citep{kurland85_optim_forag_homin_evolut}).
The main result of this work is to show and explain how evolution and collective sensing interact so that gregariousness, and hence groups, emerges from a population of randomly-initialized agents.
Initially, evolution selects agents that forage for resources for longer periods.
This individual behavior produces a pattern in the collective behavior that correlates to food location.
Agents who are able to exploit this pattern through collective sensing, i.e. gregarious agents, gain an evolutionary advantage, therefore groups form.

The contribution of this work can be summarized as follows:
(i) it unifies and expands on the results of previous work in a framework that combines evolution and social learning via neural networks \citep{hein15_evolut_distr_sensin_collec_comput_animal_popul,torney11_signal_evolut_cooper_forag_dynam_envir};
(ii) it quantifies the effect of environmental conditions on the evolutionary fitness of different foraging strategies;
(iii) it identifies environmental conditions under which a gregarious population can invade an individualistic population;
(iv) it validates previous work by showing that, in a limited region of the parameter space, gregariousness is the evolutionary-fittest strategy; and
(v) it shows empirically that groups can emerge from a randomly-initialized population as a result of interactions between natural selection and collective sensing.

The article is structured as follows. In Section \ref{lit} this work is situated within the Agent-Based Modeling literature.
The experiments are described in Section \ref{exp}, while the model is described in Section \ref{method}.
Section \ref{results} comments on the results and their implications, which are summarized and discussed in Section \ref{conclusions}.

\section{Literature review \label{lit}}

Literature on agent based modeling of group and societal issues spans over decades.
Many agent based models of society concentrate on the problem of cooperation \citep{helbing09_outbr_cooper_among_succes_driven} or coordination \citep{maes10_indiv_as_drivin_force_clust_phenom_human}.
This work considers instead a simpler scenario where agents cannot actively communicate, thus neither cooperation nor coordination are possible.
A foraging task is modeled, where agents compete for the same pool of resources.

This work relaxes many common assumptions in the literature:
\begin{itemize}
\item Environmental factors that favor groups, e.g. \citep{bowles04_evolut_stron_recip,montanier13_evolut_altruis_spatial_disper,torney11_signal_evolut_cooper_forag_dynam_envir}. Cooperation cannot be achieved as agents are unable to communicate. Moreover, agents compete for the same resources and there are no advantages in being part of a group.
\item Kin selection, e.g.  \citep{smith64_group_selec_kin_selec,hales00_cooper_memor_space,hammond06_evolut_contin_altruis_when_cooper_is_expen,helbing08_migrat_as_mechan_to_promot_cooper}. Agents do not have any visible characteristic that make them recognizable as members of a group, e.g. Green Beard \citep{dawkins76_selfis_gene} so they cannot develop mechanisms that favor kin.
\item Limited dispersal, e.g. \citep{hamilton64_genet_evolut_social_behav,nowak92_evolut_games_spatial_chaos,santos06_cooper_prevail_when_indiv_adjus,grund13_how_natur_selec_can_creat}. Agents are randomly placed and there is no mechanism that explicitly keeps offspring near their parents. Of course, this could still happen as a result of specific environmental settings, e.g. an offspring is born in a patch with food. Similarly, food is created randomly in the grid, so a new food source is not likely to be created near a depleted one.
\item Learning agents, e.g. \citep{axelrod02_beyon_geogr,nemeth07_evolut_altruis_spatial_struc_popul,duan10_fairn_emerg_from_zero_intel_agent}. Agent behavior is fully determined by their genetic characteristics.
\end{itemize}
Moreover, agents have bounded rationality \citep{simon1982models} as they have to cope with an imperfect perceptual system.
Nevertheless, the conclusions are the same as other work on collective sensing, in that collective sensing is shown to evolve from a population of individualistic agents \citep{hein15_evolut_distr_sensin_collec_comput_animal_popul}
if resources are scarce \citep{torney11_signal_evolut_cooper_forag_dynam_envir}.

  The current work reproduces previous results in a general framework, where agent behavior is determined by a neural network, that does not make any assumption about the agent's behavior.
  Previous work combined neural networks and evolutionary simulations, but the focus there was on individual learning \citep{hinton87_how_learn_can_guide_evolut,gruau1993adding, batali96_model_evolut_motiv,redko10_learn_evolut_auton_adapt_agent}.
  Another body of literature looks at interactions between neural networks, in particular the emergence of language \citep{sukhbaatar2016learning,lazaridou2016multi,foerster2016learning} and cooperation \citep{tampuu2017multiagent}, but evolution was not part of these studies.
In the proposed model, gregarious behavior is not defined by the value of a parameter but is instead a region in a $N \times M$-dimensional parameter space, where $N$ is the number of perceptions and $M$ the number of actions.
Moreover, agents are not able to emit any signal, instead evolving the capacity to respond to Inadvertent Social Information (ISI) \citep{danchin04_public_infor}.
Despite the generality of the model, gregariousness emerges as the evolutionary-fittest strategy under some environmental conditions, thus validating the previous results in a specific area of the parameter space.

\section{Experimental setting \label{exp}}
This section describes the goal and the design of the experiments, while results are presented and discussed in Section \ref{method}. Table \ref{tab:params_exp} compares how the model parameters vary across these experiments. Appendix \ref{sec:appendix} describes how to reproduce the results.

The first experiment compares the efficiency of the two competing foraging strategies, random walk and gregariousness, and investigates how it changes for different parameter configurations.
Each agent in the population is instructed to play one strategy and strategies are compared by computing the average fitness, that is, the number of successful foraging actions at the end of the simulation, of individuals implementing that strategy.
All simulations start with a population of the same size, and a fraction of it is initialized with the gregariousness strategy,
  which is obtained by initializing the weights of the neural network such that weights connecting the input and the movement action in the same direction are larger than the other weights.
  This configuration of weights ensures that at a large input in one
 direction corresponds to a large output at the corresponding movement action.
  If the input vector includes more than one large value, the winning action is determined by noise.
  The validity of this initialization is verified by means of a suit of tests,
developed with the goal of classifying agent behavior in a linear scale from random walk to gregariousness strategy \cite{burtsev06_evolut_cooper_strat_from_first_princ}.
Each test presents the agent with a predetermined perception vector and records the chosen action.
The action of foraging is tested by placing the agent in a location with food and no agents in sight, the action of moving north is tested by placing the agent in a location without food and with another agent to their north, and so on for the remaining movement actions.
All tests are performed on all agents in the population and the frequency of choosing the appropriate action is recorded: a high score in all tests means that the population behaves gregariously, while random walk scores close to 25\% on all movement actions.

The second experiment looks at the evolution of competing foraging strategies.
The experiment is set up as before, but agents reproduce with a probability proportional to their fitness, and die with a probability proportional to their age
(see Table \ref{tab:constants}).
During every reproduction, the neural weights of the offspring are initialized with a mutated version of the parent's weights.
Mutations allow the behavior to drift away from the predefined strategies and adapt over time to the environment and to the competition.
The evolution of the behavior of the two sub-populations is investigated in different environmental situations by means of the test suite described above.

The last experiment tests whether a population of randomly-initialized agents can evolve gregariousness in response to a change in the environment.
The population is initialized randomly and a strategy is allowed to evolve.
The strategy of the population is measured, as in the previous experiment, by testing the behavior of each agent at each time step.
The experiment is divided in two phases: In the first phase the resources are abundant, while in the second phase they become scarce.
The two phases are delimited by an event denominated \emph{famine}, after which the quantity of the resources decreases gradually from the high value to the low value.

\begin{table}[h]
  \centering
  \begin{tabular}[h]{l|l|l|l}
    \textbf{Parameter} & \textbf{Exp1} & \textbf{Exp2} & \textbf{Exp3} \\
    \hline
    The initial population size & [20,30,50,70,100] & 20 & 10 \\
    Number of patches containing food & [5,10,20,50,100] & 50 & 200 \\
    Field of view: radius of the perceptual system & [1,2,4,6,9] & 1 & 1\\
    Length of simulation & 20k & 50k & 50k \\
    The initial fraction of gregarious agents & 0.5 & 0.1 & - \\
  \end{tabular}
  \caption{Comparison of parameter values across experiments.}
  \label{tab:params_exp}
\end{table}

\section{Methods \label{method}}
\label{sec-3}
The effect of collective sensing in group formation is investigated by means of an agent-based evolutionary model \citep{epstein99_agent_based_comput_model_gener_social_scien}.
Simulation is a suitable tool, and preferable over mathematical models for studying processes generated by individual interactions, as it allows for capturing their dynamics \citep{healy67_lanal_mathem_des_faits_sociaux}.

A population of agents performs a foraging task in a patchy environment \citep{code}, which is modeled on previous work \citep{beauchamp00_learn_rules_social_forag,hamblin09_findin_evolut_stabl_learn_rule}.
The environment is a square grid with periodic boundary conditions in which a fixed number of patches contain a random positive quantity of food units.
Food quantity is sampled from a uniform distribution with a mean of 100. This allows several agents to forage from the same resource for a short time.
The total number of food sources is governed by a model parameter: whenever a food source is exhausted, a new one is spawned at a random location.

\begin{figure}[h]
\centering
\includegraphics[width=.9\linewidth]{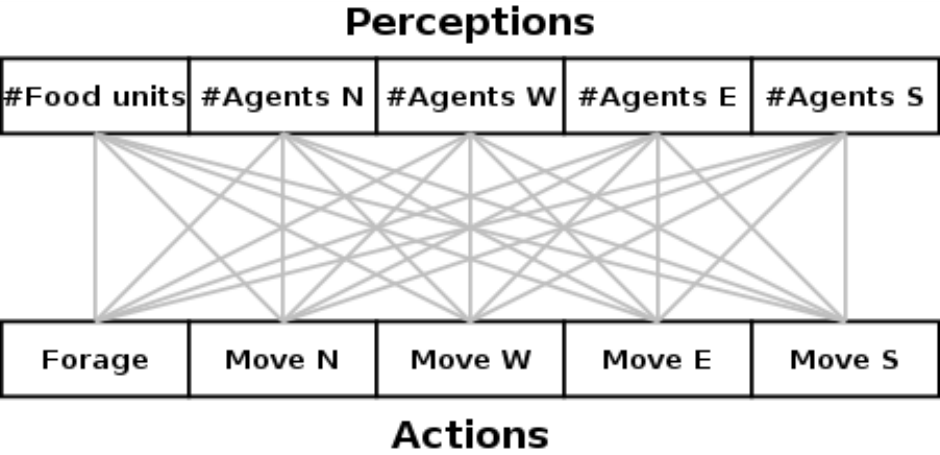}
\caption{\label{fig:nn}Diagram of the neural network driving the agent behavior. Each link between a perception and an action represents a weight.}
\end{figure}

Agent behavior is driven by a neural network which connects each perceptual input to each of the possible actions (see Figure \ref{fig:nn}).
  Given a perception vector, the score of each action is the sum of all input values, multiplied by the corresponding weight that connects each input to the action, and by random noise.
  The action with the highest score is executed. If several actions obtain a similar score, the choice of one action over another is determined by a combination of noise, small differences in the input values and/or small differences in the weights.
  Behavior is selected by the evolutionary process among a vast set of alternatives, i.e. any of all possible mappings from any input vector to an action.
  The use of neural networks as controllers removes the need for modeling assumptions, e.g. a social parameter, which might limit the space of the possible behaviors.
Agents are initially placed at random locations at the start of the experiment.
At every timestep agents play their turn one after the other, in a random order: each agent perceives the environment and executes one action.
Randomization of the order of play is important as agents compete for the same rivalrous resources, i.e., food consumed by an agent is not available for other agents.
Agents' actions are either those of movement or of foraging: Agents can spend an action moving of one patch in any direction, alternatively each agent can forage one unit of food from the current patch.
Foraging is successful if the current patch contains at least one food unit, which is then consumed and removed from the patch.
Foraging from a patch that does not contain any food produces a foraging failure.
Fitness is defined as the number of successful foraging actions during an agent's lifetime, thus fitness measures the quality of the individual foraging strategy.
For simplicity, we assume that energy is never consumed, thus fitness is equivalent to energy.

Agents reproduce with a probability proportional to their energy.
Reproduction spawns a new agent in the same patch, and energy is equally split between parent and offspring.
  The offspring's behavior is determined by a mutated version of the parent's weights. Mutations increase or decrease the original weights by a small random value.
Agents are removed from the game with a probability proportional to their age.
Both reproduction and death are modeled with a roulette wheel algorithm with stochastic acceptance (as in \citep{torney11_signal_evolut_cooper_forag_dynam_envir}).

Agents perceive the quantity of food and agents present in their surroundings.
The perception of food is limited to only the current patch, while agents can be perceived within a given distance.
This assumption is often verified in nature \citep{kurland73_natur_histor_kra_macaq_fascic,klein73_obser_two_types_neotr_primat_inter_assoc,kurland85_optim_forag_homin_evolut} and is also a prerequisite for collective sensing.
The perception mechanism does not distinguish the number of agents in each visible patch, it aggregates instead the number of agents by angular direction \citep{strandburg-peshkin13_visual_sensor_networ_effec_infor}.
For simplicity, the field of view is subdivided into four areas corresponding to the cardinal directions and a parameter called ``field of view'' determines how far agents can see: a higher value implies that each of the four areas contains more patches, therefore perception is more coarse-grained.
A higher number of agents in one area corresponds to a higher input signal: the more crowded the surrounding patches are, the higher the input signals, thus if the population grows large, as in the evolutionary simulations, it becomes more difficult to perceive the signal carried by collective sensing.
  For this reason, it is assumed that the perceptual system can distinguish whether an agent is foraging or moving \citep{10.3389/fncir.2014.00119}, and that agents are considered only if they are foraging \citep{bhattacharya14_collec_forag_heter_lands}. This allows the supporting of larger populations in the environment.
  This assumption does not change the results qualitatively: Figure \ref{fig:normperc} confirms that gregarious agents have higher fitness than random walkers, also when this assumption is relaxed \citep{bennati16_role_infor_group_format}.

Collective sensing allows individuals to perceive an invisible signal (food location) by means of a visible proxy signal (group dynamics), thus the proxy signal must be visible where the other signal is not.
In this case the proxy signal is the location of other agents, as a higher concentration of agents correlates to the presence of a food source, under the assumption that agents remain on food sources (cf. position-dependent diffusion \citep{schnitzer93_theor_contin_random_walks_applic_to_chemot})
Interestingly, this signal does not correlate with food availability. In fact the stronger the signal the more agents there will be foraging in a patch and the faster the food resource gets depleted.

\begin{figure}[t!]
  \centering
  \begin{minipage}[t]{\linewidth}
  \centering
\includegraphics[width=0.45\textwidth]{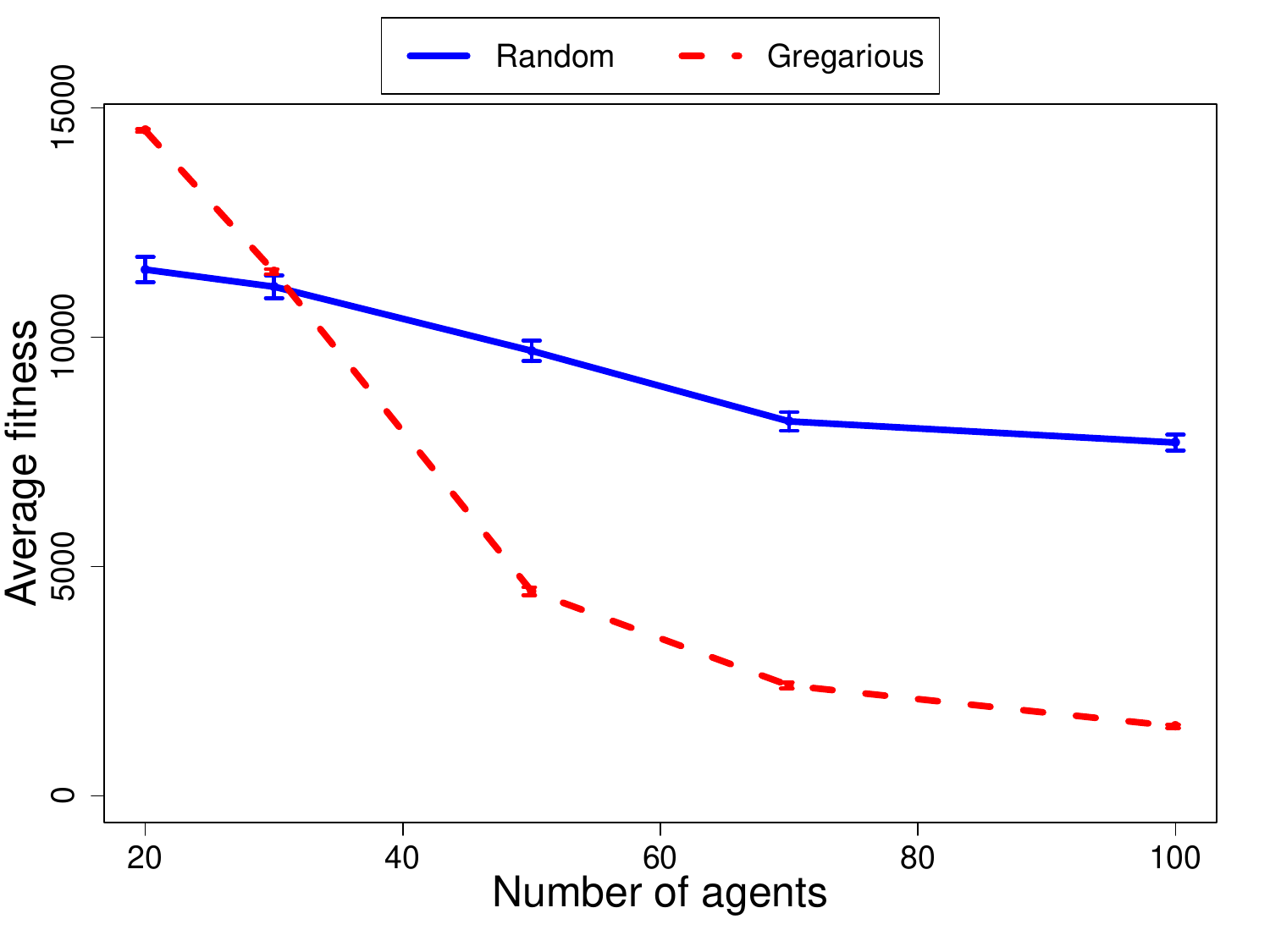}
\includegraphics[width=0.45\textwidth]{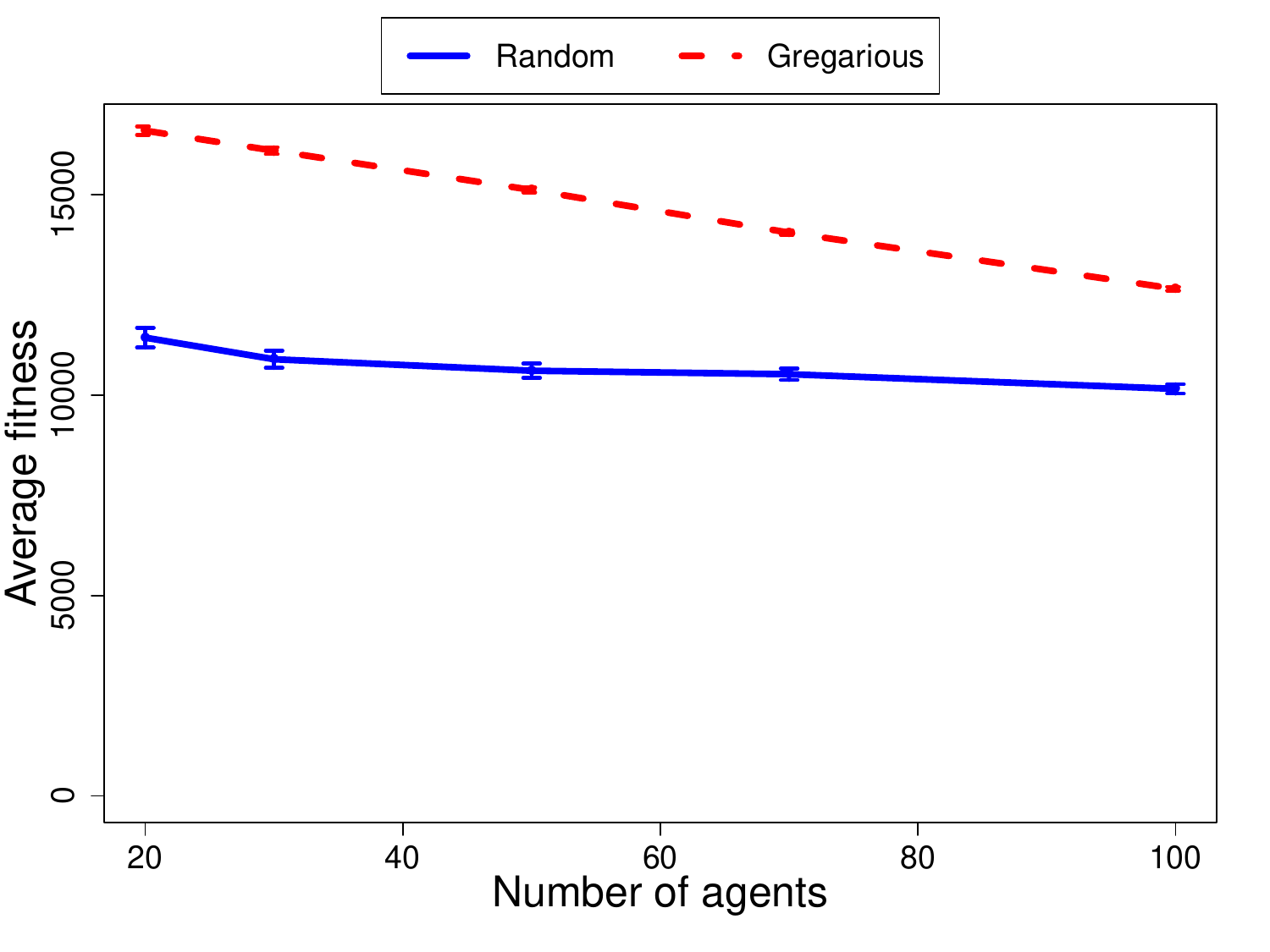}
\caption{  \label{fig:n}Average fitness for increasing population size. Initial size of the gregarious subpopulation: 50\% (left) and 10\% (right). Error bars represent 0.95 confidence intervals. Parameters of the simulation: Field of view 4 and 50 food sources.}
  \end{minipage}
  \begin{minipage}[t]{\linewidth}
  \centering
\includegraphics[width=0.45\textwidth]{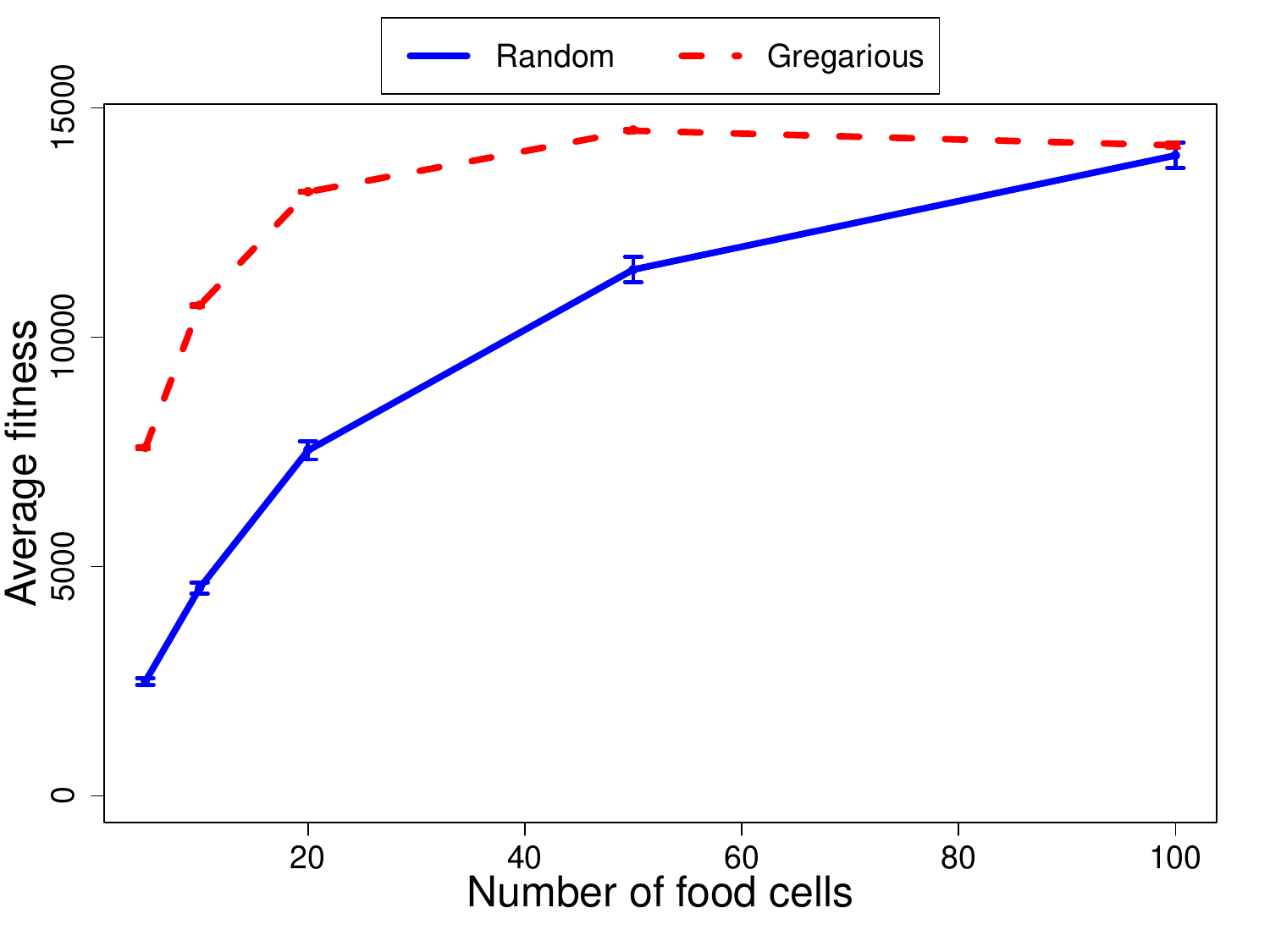}
\includegraphics[width=0.45\textwidth]{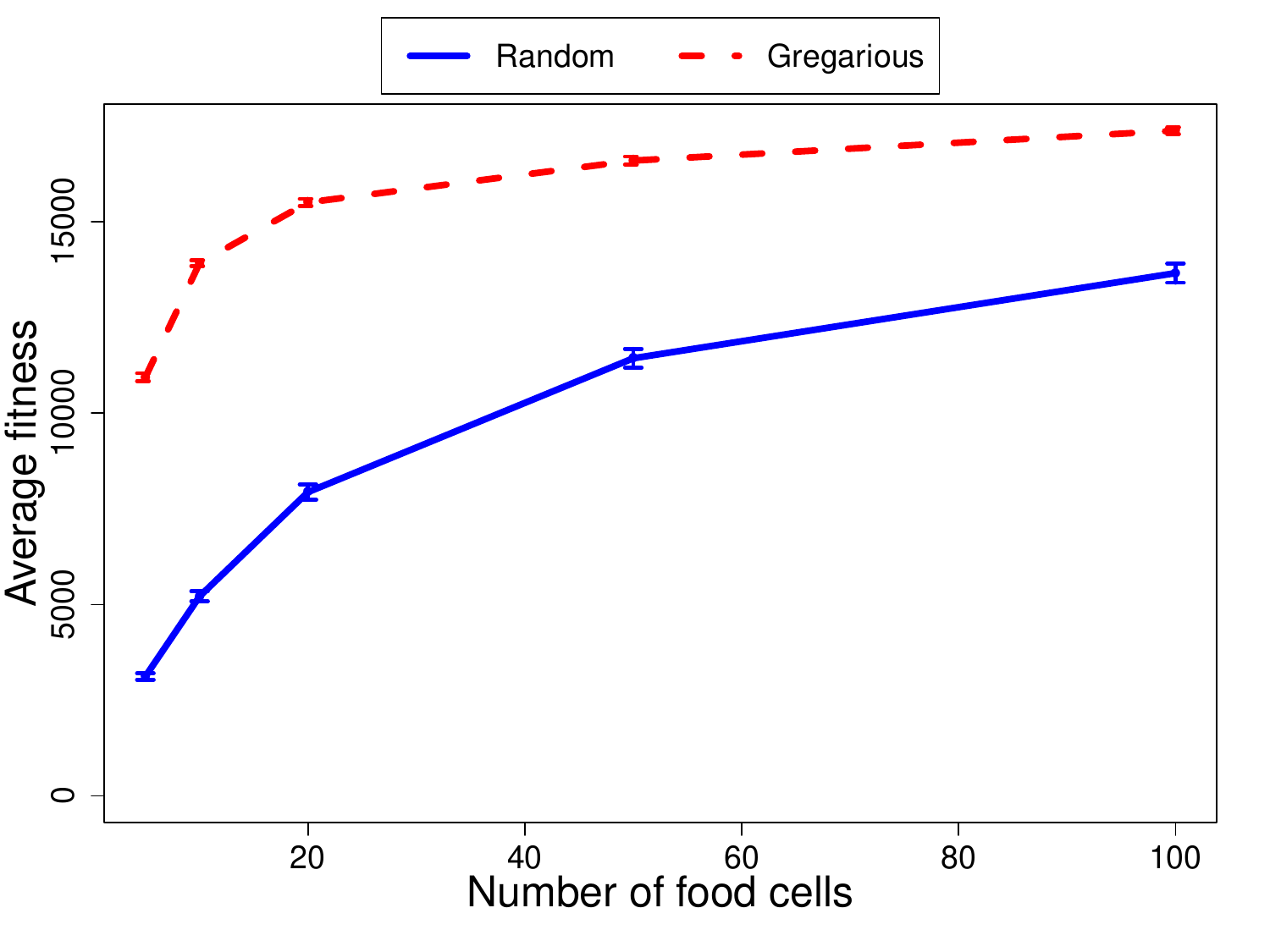}
\caption{\label{fig:nf}Average fitness for increasing number of resources. Initial size of the gregarious subpopulation: 50\% (left) and 10\% (right). Error bars represent 0.95 confidence intervals. Parameters of the simulation: Population of size 20 and field of view 4.}
  \end{minipage}
  \begin{minipage}[t]{\linewidth}
  \centering
\includegraphics[width=0.45\textwidth]{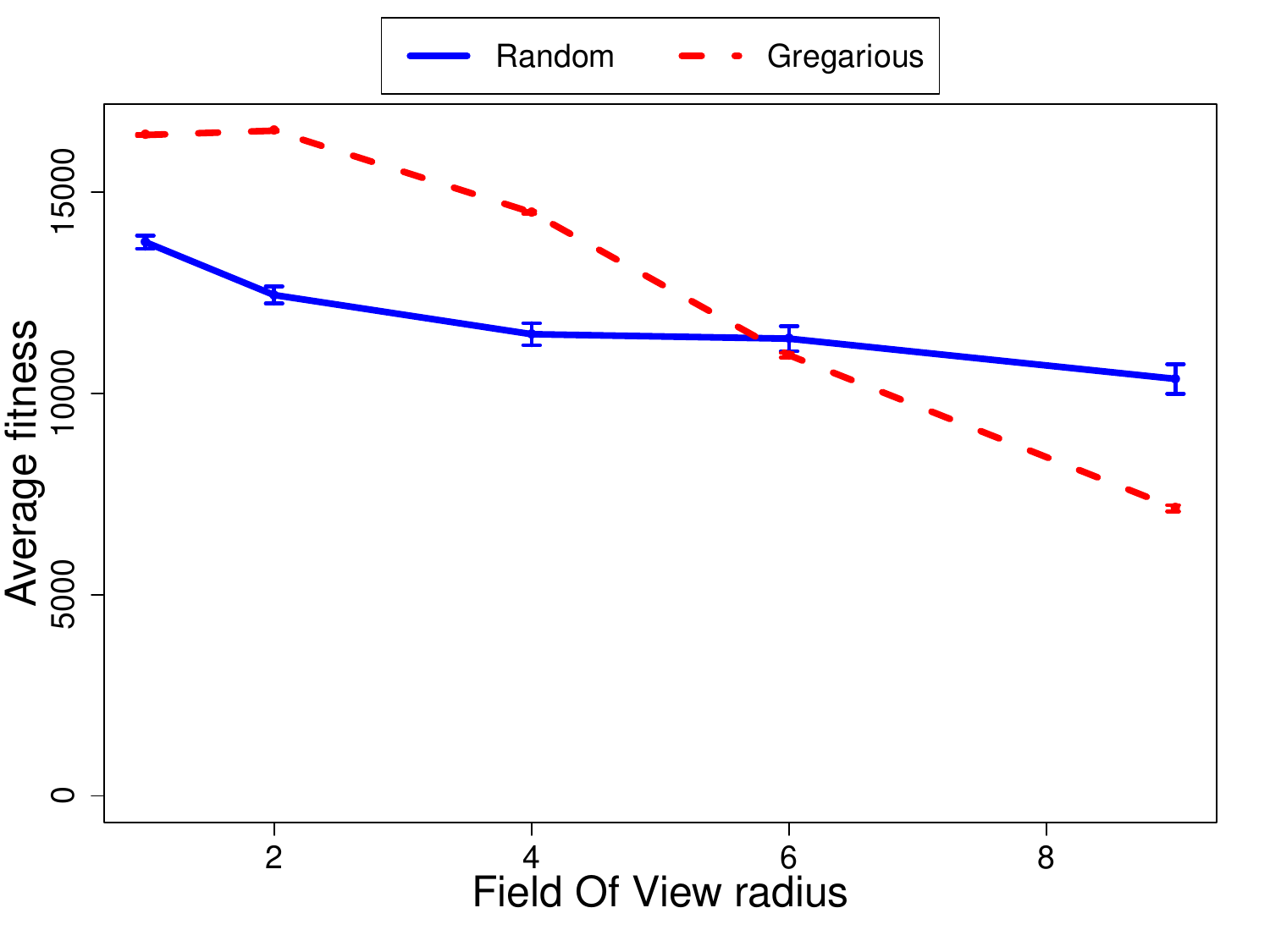}
\includegraphics[width=0.45\textwidth]{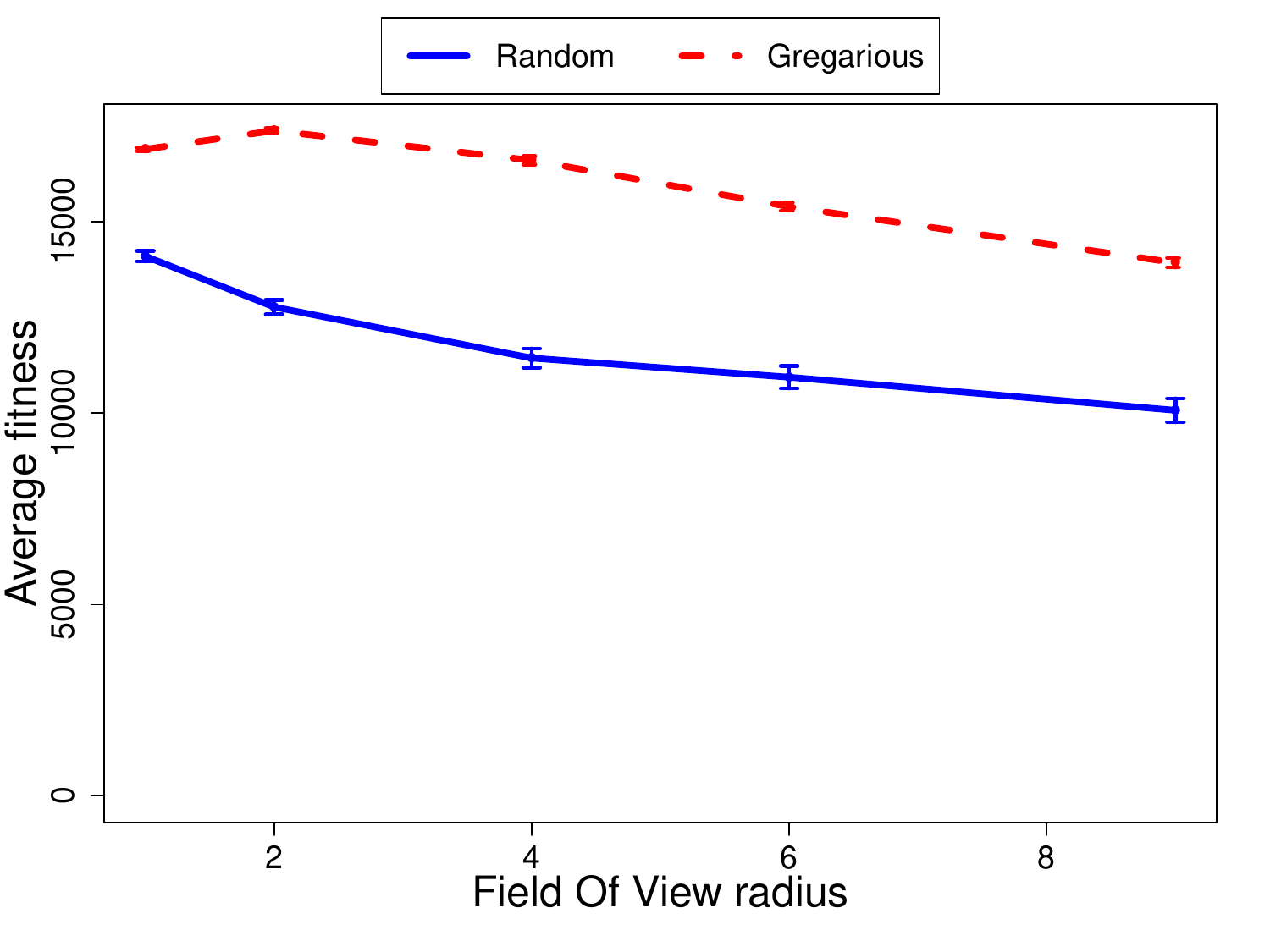}
\caption{\label{fig:fov}Average fitness for increasing field of view. Initial size of the gregarious subpopulation: 50\% (left) and 10\% (right). Error bars represent 0.95 confidence intervals. Parameters of the simulation: Population of size 20 and 50 food sources.}
  \end{minipage}
\end{figure}

\section{Results \label{results}}
\label{sec-5}
In the first experiment two foraging strategies, gregariousness and random walk, are compared.
The fitness of a strategy is measured by the average number of resource units foraged by all agents implementing such a strategy.
The main result is that gregariousness can be a more efficient strategy than random walk.
If food sources are rare, gregarious agents have a higher evolutionary fitness than random walkers, which decreases with population size (see Figure \ref{fig:n}) as there is higher competition for resources.
    Abundance of food reduces the advantage of gregarious strategy over random walk, until it eventually disappears (see Figure \ref{fig:nf}).

The gregariousness strategy allows agents to exploit information about the behavior of other agents to find sources of food.
Whenever an agent finds a food source and exploits it, gregarious agents start converging and eventually exploit the same resource.
This behavior leads agents to concentrate on food locations, thus the fewer the food locations the more crowded they are.
Large groups of agents in one patch produce a strong signal, which gregarious agents can exploit to their advantage.
Gregarious agents compete with each other for the same resources, while random walkers explore the environment for new resources, therefore a larger number of gregarious agent increases competition hence reducing their average fitness (see Figure \ref{fig:n})
After an agent finds a new resource, it can exploit it exclusively until another agent reaches the same position; the longer an agent can exploit a resource exclusively, the higher the benefit from exploring the environment.
Therefore exploration, i.e. random walk, becomes advantageous if food is abundant because the probability of randomly finding a new food resource increases.

The effect of the field of view of agents on their performance is shown in Figure \ref{fig:fov}.
Gregarious agents see their fitness decrease with an increasing field of view, an effect of the perception system's design which aggregates the contents of multiple cells in a region.
The more agents are in the field of view, the more difficult it becomes to distinguish whether the region contains a crowded location or many locations with few agents.
This effect can be reduced with a more sophisticated perception system which is able to distinguish counts at the patch level.

\begin{figure}[h!]
  \centering
\includegraphics[width=0.5\textwidth]{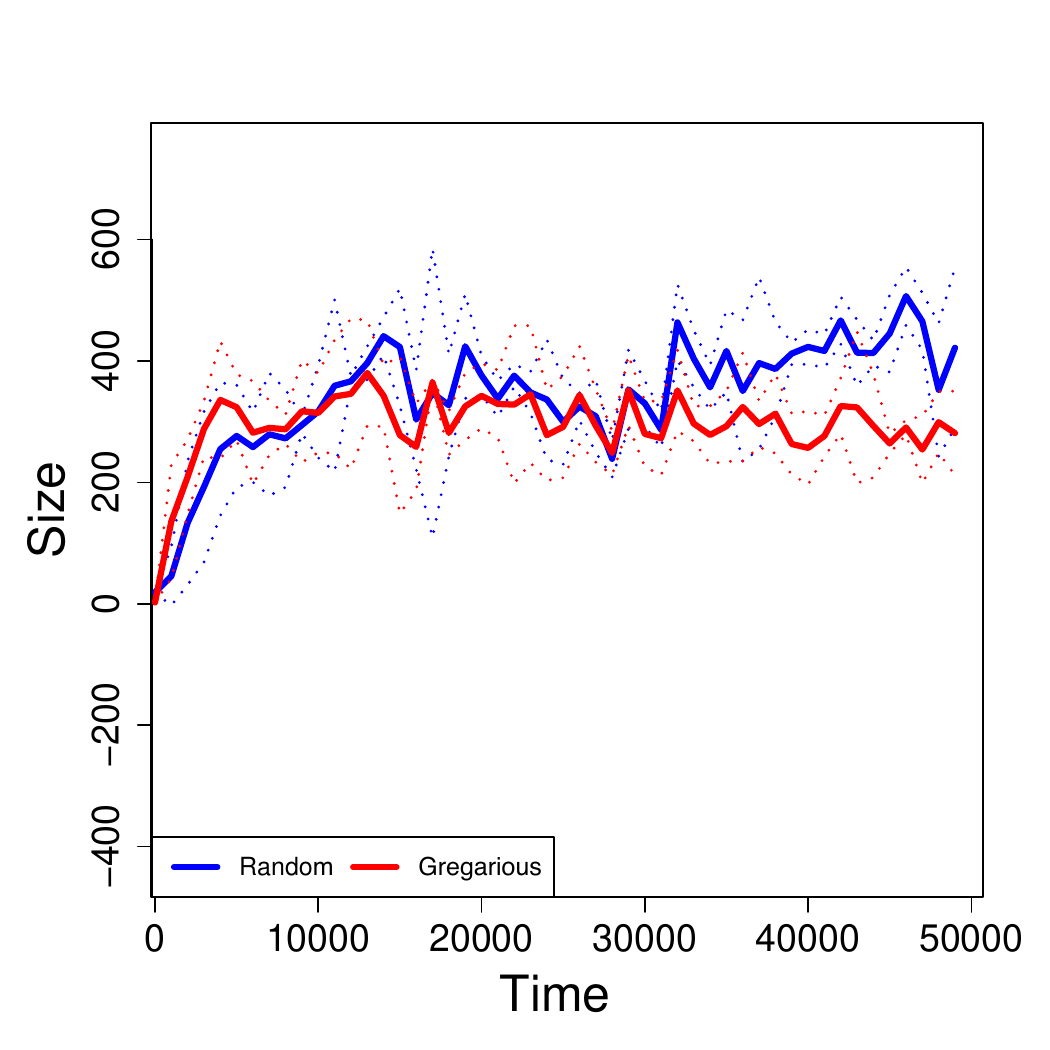}
\caption{ \label{fig:evo}Average population size over time. Dotted lines represent 0.95 confidence intervals. Parameters of the simulation: Field of view 1 and 20 food sources.}
\end{figure}
\begin{figure}[h!]
  \centering
  \includegraphics[width=0.45\textwidth]{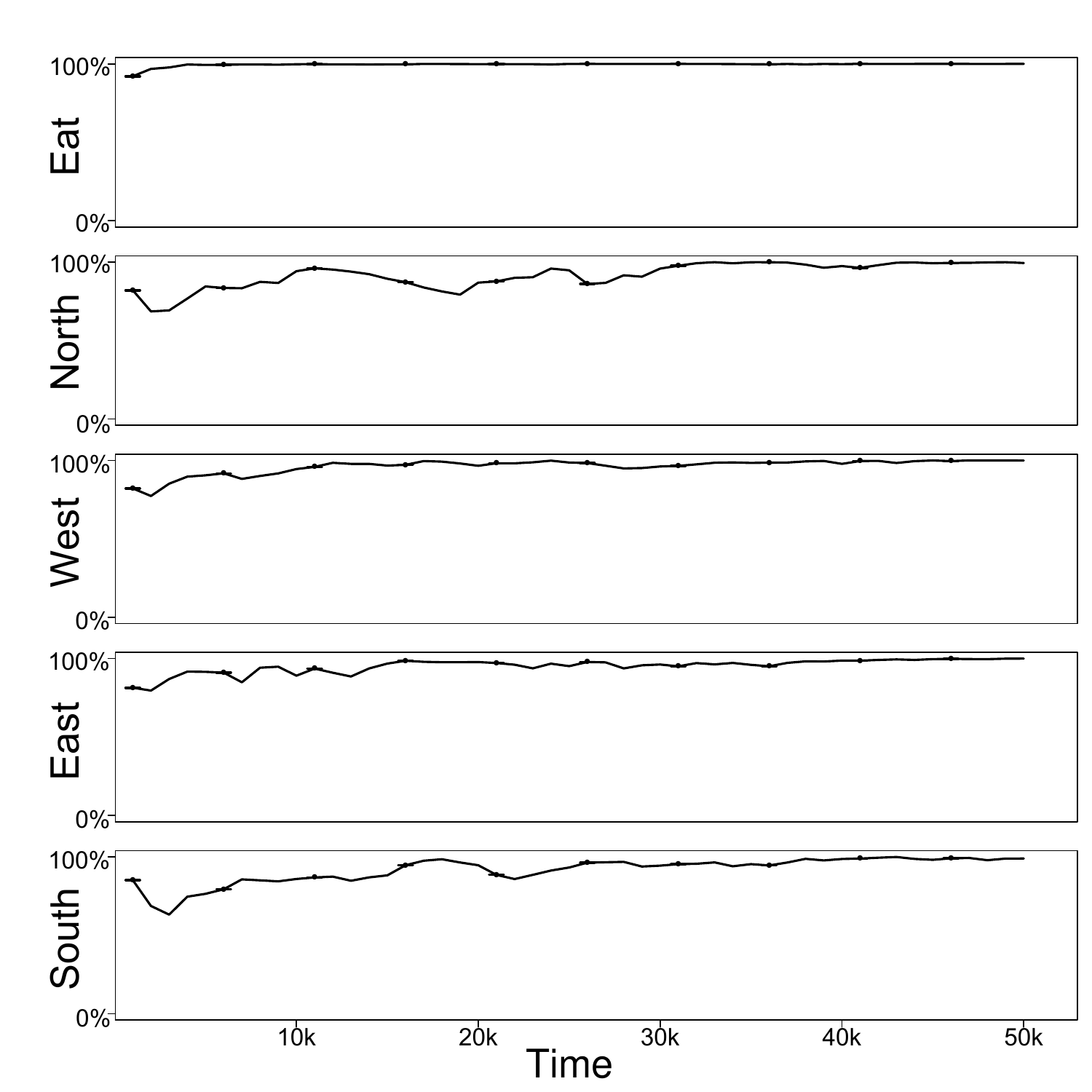}
  \includegraphics[width=0.45\textwidth]{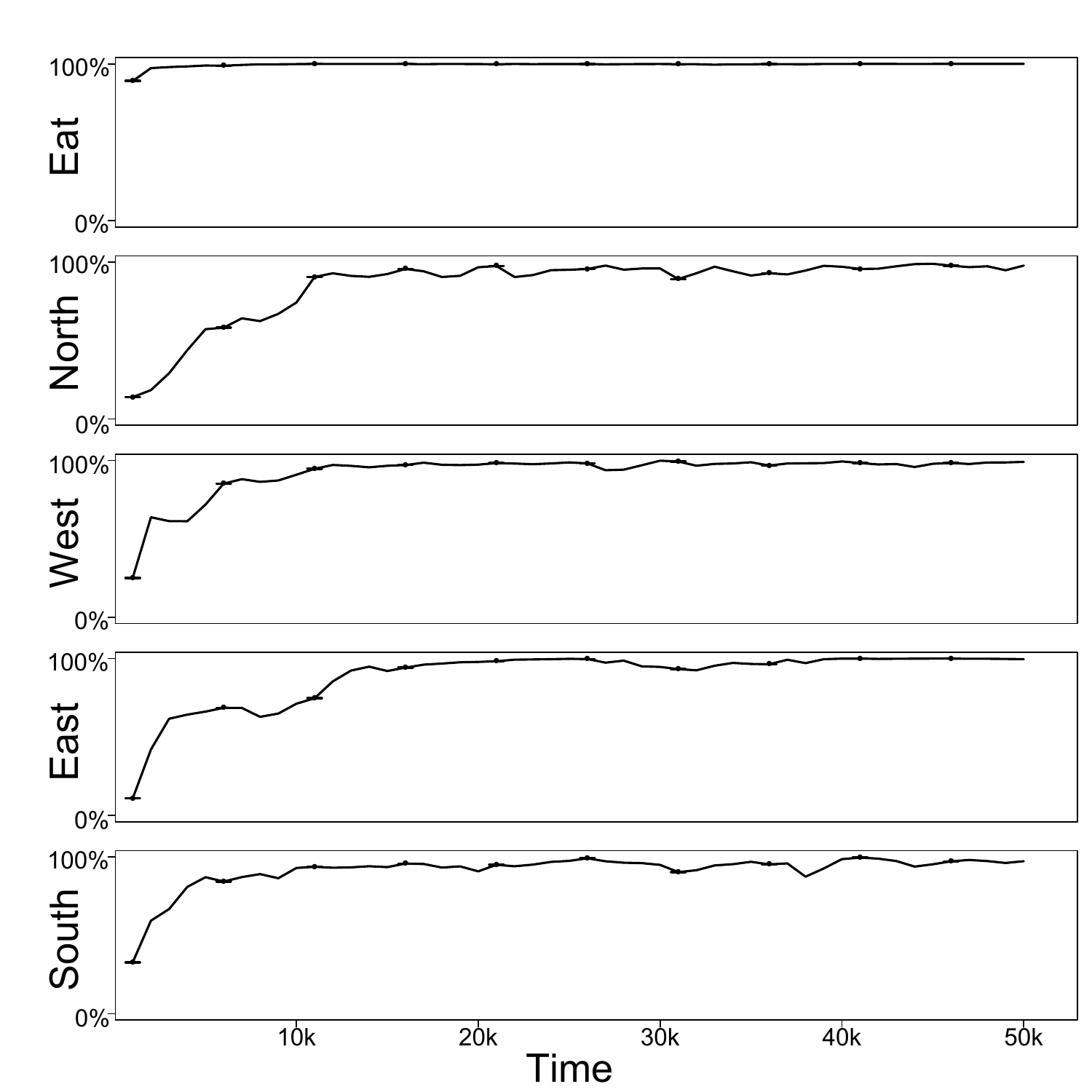}
  \caption{Evolution of strategy over time. Each line represents the frequency at which one action is executed according to the gregarious strategy: for example a value of 80\% for action 'left' indicates that action 'left' is executed 80\% of the time in a situation where gregarious strategy would dictate 'left'. Left: Gregarious agents keep their behavior consistent with the gregarious strategy, Right: random walkers start with a random walk (each movement action is executed approximately 25\% of the time) but rapidly adopt a gregarious behavior. Error bars represent 0.95 confidence intervals. Parameters of the simulation: Initial population of size 20 and 50 food sources, field of view 1.}
  \label{fig:evo:decisions}
\end{figure}

The second experiment repeats the comparison of two foraging strategies in an evolutionary setting, where agents can die and reproduce.
  Figure \ref{fig:evo} shows the change in size of the two groups over the course of the simulation.
This effect is explained by the evolution of the individual behaviors:
  gregarious agents keep their behavior constant for the entire simulation (see Figure \ref{fig:evo:decisions} left), while random walkers adapt their behavior until they adopt the gregarious strategy (see Figure \ref{fig:evo:decisions} right).
  The whole population adopts the gregarious strategy at around time 10000, which confirms that gregariousness is the evolutionary-fittest strategy.

\begin{figure}[h]
\centering
\includegraphics[width=.7\linewidth]{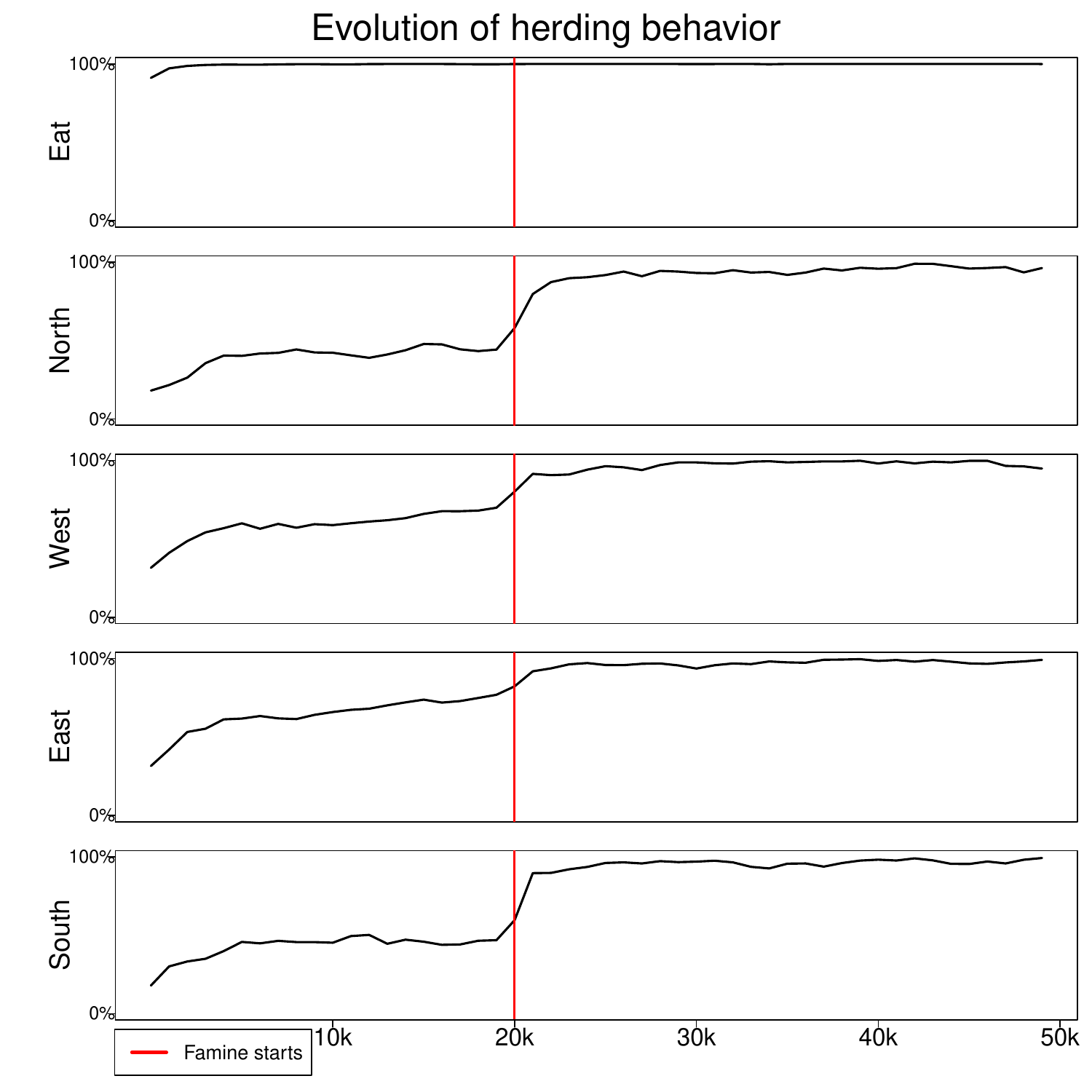}
\caption{\label{fig:famine}Strategy evolution of random walkers over time. Average individual behavior over 24 different population. Parameters of the simulation: Initial population of size 10 and 150 food sources. At timestep 20000 the number of food resources is reduced to 5.}
\end{figure}

\begin{figure}[h]
\centering
\includegraphics[width=0.8\textwidth]{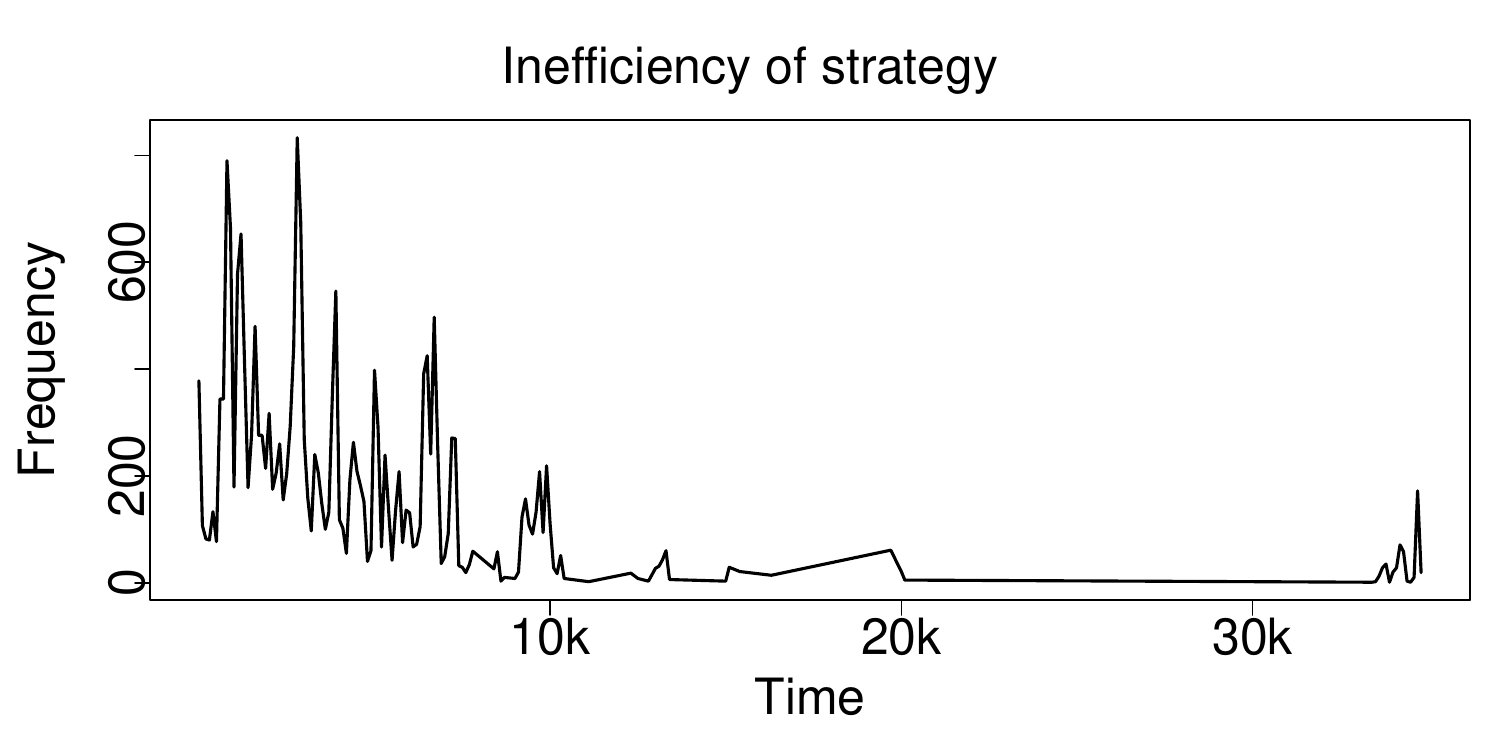}
\caption{\label{fig:pointless}Inefficiency of foraging strategy. The graph shows the change over time in the frequency of inefficient actions, i.e. foraging failures, performed by the agents.}
\end{figure}

In the third experiment a population of randomly-initialized agents is let evolve in a variable environment.
The strategy adopted by the population is evaluated by testing the behavior of each agent at every timestep (see Figure \ref{fig:famine}).
The simulation starts with abundant resources; agents are quickly selected on the basis of their ability to forage but their movement is largely independent from the perceptual stimuli.
At one point during the simulation, signaled by the red vertical line, the number of resources drops and the population rapidly adopts gregarious behavior in response to this change in the environment.
One possible interpretation for this result is that some agents learn gregarious behavior at the beginning of the simulation, but the selective pressure is not enough to remove other less efficient strategies from the population until resources become scarce.
Another interpretation is that selective pressure is high enough to remove inefficient strategies from the start, therefore random walk is indeed the most efficient strategy when the resources are abundant.
In order to distinguish between these two interpretations, it is necessary to measure the efficiency of a strategy; the frequency of foraging failures is chosen as a measure for strategy efficiency.
The frequency of foraging failures approaches zero after 10,000 timesteps, which means that efficient strategies are selected at the beginning, supporting the second interpretation (see Figure \ref{fig:pointless}).
This implies that, in a collective sensing scenario, gregariousness is the evolutionary-fittest strategy when food sources are rare, while random walk is the fittest strategy otherwise.

This result validates previous work on collective sensing by showing that groups do indeed emerge from a population of individualistic agents without any assumption about their behavior, but only if resources are scarce.
Moreover, it shows that evolution favors from the start strategies that produce a low number of foraging failures, thus creating a correlation between the location of agents and resources.
Without evolution selecting for efficient strategies, collective sensing would not be able to track the location of food, therefore gregarious behavior would not provide an evolutionary advantage.

\section{Conclusion \label{conclusions}}
\label{sec-6}
This paper studies the role of collective sensing in the formation of groups in a population of individualistic agents.
An agent-based model of a foraging task is used to study the evolution of foraging strategies in a patchy environment with limited resources.
Agent behavior is determined by a neural network which translates perceptual inputs into actions, without any explicit model about the environment and the interactions between agents.
The generality of results is increased by the relaxation of many common assumptions in the literature, thanks to a simple environment and the use of neural networks.

Two different foraging strategies are compared: random walk and gregarious behavior, i.e. moving towards other agents.
Results show that gregarious behavior provides an evolutionary advantage at the individual level through collective sensing at the group level.
Depending on the characteristics of the environment, and specifically in the case of scarce resources, this advantage can overcome the competition caused by sharing resources and groups can spontaneously form in a population of individualistic agents.
Moreover, it is shown that the evolutionary-fittest strategy is random walk if resources are abundant and gregarious behavior if resources are scarce.
This result, caused by the interaction between evolution and collective sensing, validates and unifies previous work on collective sensing under a general framework that combines evolution and social learning via neural networks.

  The framework introduced in this paper is relevant to the literature about collective sensing as it reproduces results from different models, without requiring the assumption of a 'social' parameter.
  A region in the space of environmental parameters is identified that invalidates these results, thus calling for further investigation about the effect of the environment on the previous models.
  A mechanism is individuated, i.e. there is a sudden decrease of resource availability  which could have triggered the creation of gregarious behavior in a population of solitary individuals.
  The results presented in this paper may be of interest to the fields of Biology and Social Sciences, as they introduce collective sensing as another mechanism that supports the existence of groups.
  Finally, a similar effect of resource availability on the efficiency of gregarious behavior could appear in other scenarios in which gregarious behavior is present, such as financial markets \citep{devenow1996rational}, product choice \citep{huang2006herding} and forecasts \citep{cote1997herding}.

\section{Acknowledgments}
\label{sec-7}
The author acknowledges support by the European Commission through the ERC Advanced Investigator Grant 'Momentum' (Grant No. 324247).

The author thanks Leonard Wossnig and Johannes Thiele for their help with writing the simulation software, Andrew Berdahl and Dirk Helbing for many useful comments, and several anonymous reviewer for their constructive criticism.

\appendix
\section*{Appendix}
\section{Reproducibility}\label{sec:appendix}
The datasets generated during and/or analysed during the current study are available from the corresponding author on reasonable request.
The source code used to generate and analyze the datasets is available on GitHub. \citep{code}.

A C++ compiler with MPI support is required in order to compile the code. The code has been compiled with Make and the GCC compiler, but other development environments might be compatible as well.
Data analysis and figures are produced by R, the code relies on the executable Rscript to run the analysis non-interactively.
Compilation and startup scripts are written for bash on a *nix system, but other shells might be supported as well.
The code has support for the LSF platform for parallel execution on clusters, but it can also be run on a single machine.

In order to run the code, execute the script ``build.sh'', which will build the appropriate code for each simulation scenario and start the simulation.
Parameters for each scenario are found in the ``params.sh'' file in the respective folder.
After the simulation has completed, the analysis scripts, ``analysis.R'' and ``time\_series\_3d.R'', are executed. Figures are produced for each simulation in the subfolder ``results''.
The output of the simulation might take large amounts of space on disk, therefore files are zipped after the analysis is completed. The scripts rely on ``gzip'' for compression.

Experiments differ in the parameters, contained in the file ``params.sh'', and in the features of the simulation, encoded as compile flags in the Makefile.

\begin{table}[h]
  \centering
  \begin{tabular}[h]{r|l}
  \textbf{Parameter} & \textbf{Description} \\
  num\_runs & number of timesteps \\
  samples & number of repetitions \\
  sizes & initial population size \\
  srs & proportion of gregarious agents \\
  ages & determines max\_age (cfr. \textit{pd} in Table \ref{tab:constants}) \\
  fovs & range of field of view \\
    foods & number of cells with food \\
  \end{tabular}
  \caption{Description of simulation parameters.}
  \label{tab:param}
\end{table}

\begin{table}[h]
  \centering
  \begin{tabular}[h]{r|l|l}
      \textbf{Constant} & \textbf{Description} & \textbf{Default}\\
    amin & minimum population size & 50 \\
    fmax & maximum food in each cell & 200\\
    samples & number of independent simulations & 50 \\
  decision\_noise & noise on choosing the action & 0.1\\
  weight\_noise & noise on mutating the genome & 3 \\
    max\_weight & determines the range of weights & $\pm 100$ \\
    field\_size & the size of the squared grid & 20 \\
    pr & energy/pr determines the probability of reproduction & 700 \\
    pd & age/pd determines the probability of death & 1000 \\
  \end{tabular}
  \caption{Description of constants in the simulation.}
  \label{tab:constants}
\end{table}

\begin{table}[h]
  \centering
  \begin{tabular}[h]{r|l}
    \textbf{Flag} & \textbf{Description} \\
  \hline
    debug & activates debug prints \\
    famine\_iteration & timestep at which the famine starts \\
    invisible\_food & food cannot be seen at a distance \\
    interact & agents can see each other \\
    filter\_static\_perceptions & roaming agents are not visible \\
    seed\_foraging & ``teaches'' agents to forage when there is food \\
    immortals & disables evolutionary process (birth and death) \\
    respawn & generates a new population if all agents are dead \\
  \end{tabular}
  \caption{Description of compile flags.}
  \label{tab:flags}
\end{table}

\begin{figure}[h]
  \centering
  \begin{minipage}{1.0\linewidth}
      \centering
    \includegraphics[width=0.6\textwidth]{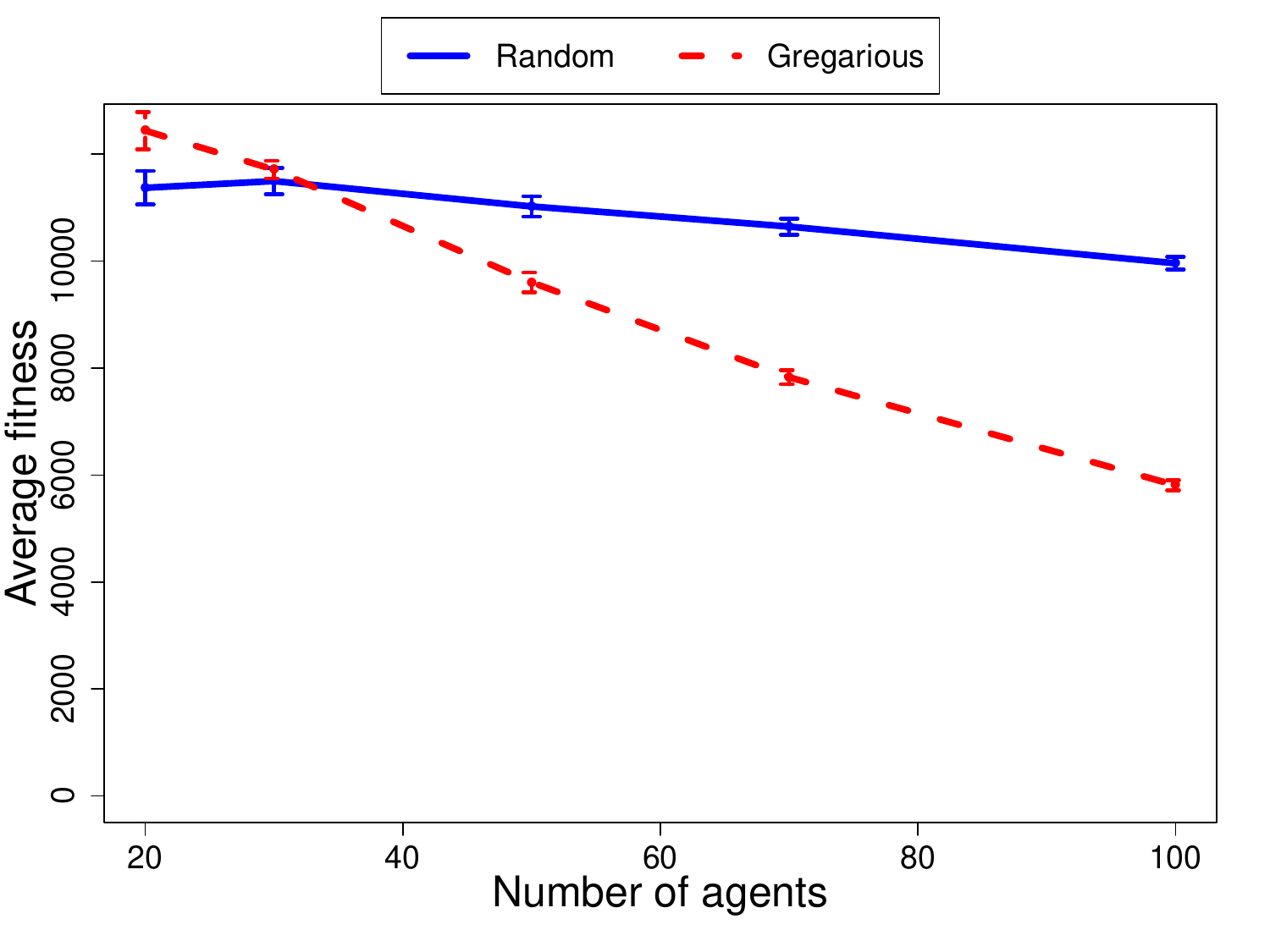}
    \subcaption{Parameters of the simulation: Field of view 4, 20 food sources, 10\% gregarious agents .}
  \end{minipage}
  \begin{minipage}{1.0\linewidth}
      \centering
    \includegraphics[width=0.6\textwidth]{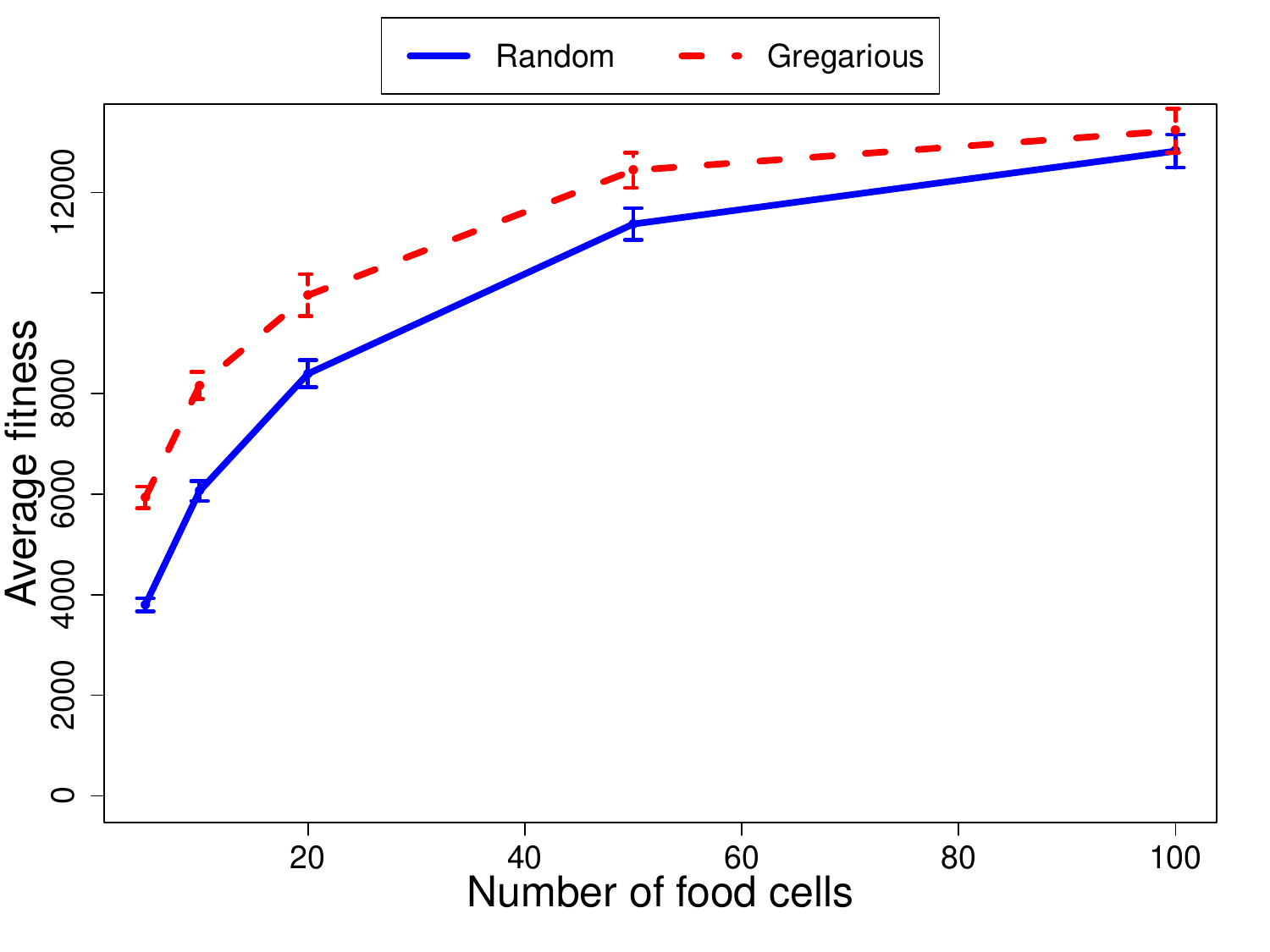}
    \subcaption{Parameters of the simulation: Field of view 4, population size 20, 10\% gregarious agents .}
  \end{minipage}
  \begin{minipage}{1.0\linewidth}
      \centering
    \includegraphics[width=0.6\textwidth]{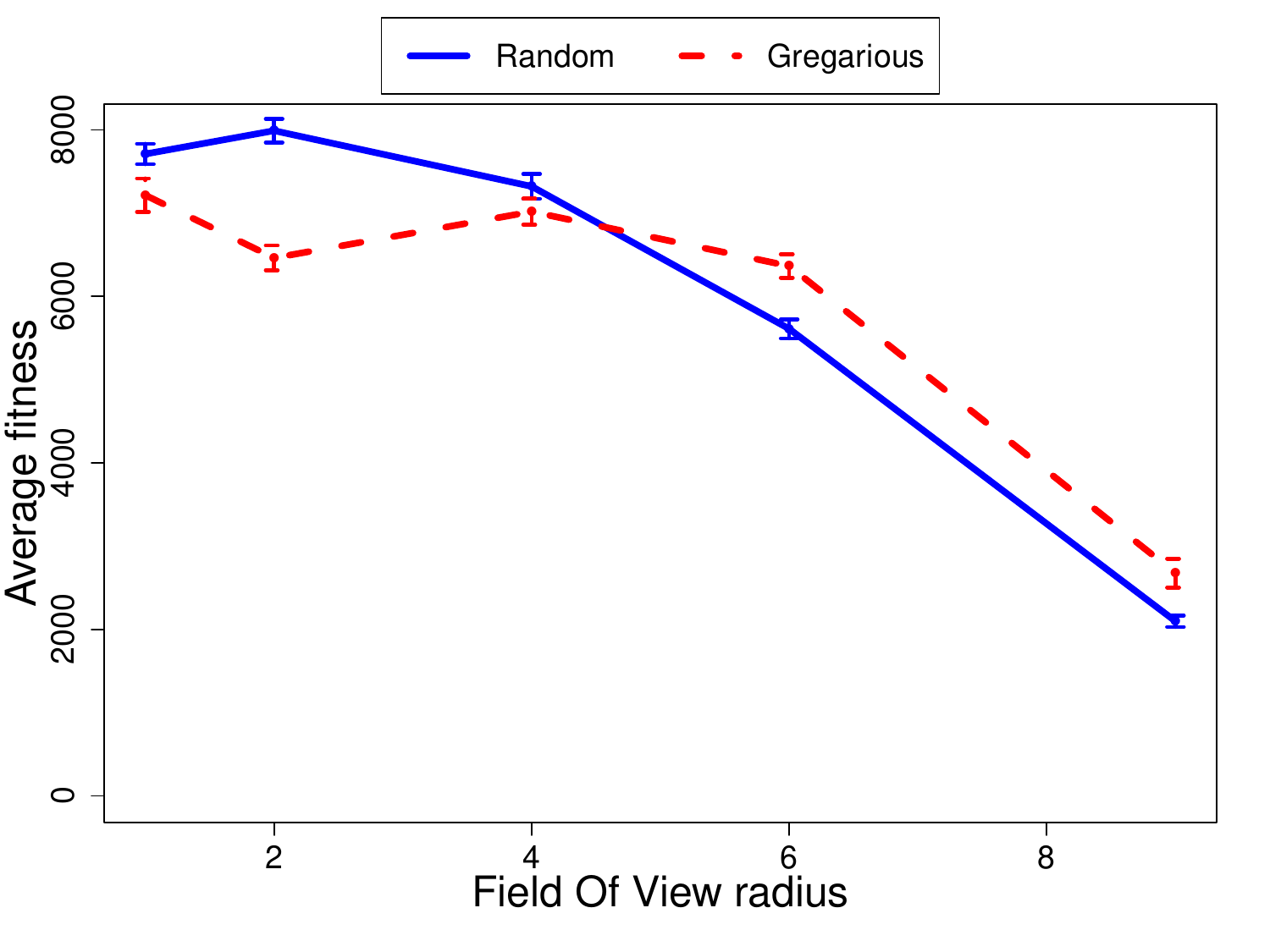}
    \subcaption{Parameters of the simulation: Population size 20, 20 food sources, 10\% gregarious agents .}
  \end{minipage}
\caption{ \label{fig:normperc}Gregarious agents have an evolutionary advantage also when relaxing the assumption in their perception.}
\end{figure}
\end{document}